\documentclass[aps,amsmath,amssymb,twocolumn]{revtex4-2}

\usepackage[]{graphicx}
\usepackage{color}
\usepackage{bbold}
\usepackage{braket}
\usepackage[colorlinks=true, pdfstartview=FitV, linkcolor=blue, citecolor=red, urlcolor=black]{hyperref}
\newcommand{\be}{\begin{equation}}
\newcommand{\ee}{\end{equation}}
\newcommand{\ben}{\begin{eqnarray}}
\newcommand{\een}{\end{eqnarray}}
\newcommand{\bes}{\begin{subequations}}
\newcommand{\ees}{\end{subequations}}
\def\bal#1\eal{\begin{align}#1\end{align}}

\newcommand{\nn}{\nonumber\\}
\newcommand{\bfi}{\begin{figure}}
\newcommand{\efi}{\end{figure}}
\newcommand{\bc}{\begin{center}}
\newcommand{\ec}{\end{center}}
\newcommand{\sech}{\mbox{sech}}
\newcommand{\sgn}{\mbox{sgn}}

\begin{document}
\title{Thick brane models in generalized mimetic gravity}
\author{A. S. Lob\~ao Jr.}
\affiliation{Escola T\'ecnica de Sa\'ude de Cajazeiras, Universidade Federal de Campina Grande, 58900-000 Cajazeiras, PB, Brazil}
\begin{abstract}
In this work, we investigate braneworld models in mimetic gravity, with the source field having generalized dynamics. We present the mathematical description necessary to study the generalized model and consider the first-order formalism to solve the equations of motion. We also investigate the linear stability of the gravitational sector to verify that the model is stable. In particular, we consider two specific situations where we were able to get exact solutions with a kink-like profile and calculate several important properties of the brane.
\end{abstract}

\maketitle
{\it {1. Introduction.}} Braneworld models are well-established theories of gravity with extra dimensions that emerged in the late 1990s as a theoretical proposal to explain the hierarchy problem \cite{Randall:1999vf}. In the original scenario, the brane was thin, however, the inclusion of scalar fields as sources of the brane made the brane thick and allowed the incorporation of new and interesting possibilities, for example, the emergence of internal structures \cite{Goldberger:1999uk, Skenderis:1999mm, DeWolfe:1999cp, Csaki:2000fc,Brito:2001hd}. Nowadays, we may say that the study of gravity in higher dimensions has a solid theoretical basis, still being a prominent field of investigation.

Recently, much attention has been paid to the study of braneworld models in modified gravity, see \cite{Rosa:2020uli,Bazeia:2020jma,Fu:2021rgu,Rosa:2021tei,Moreira:2021cta,Silva:2022pfd,Bazeia:2021bwg} for some recent work. In general, the proposals in this line attempt to reconstruct generalized four-dimensional gravity models to minimally reproduce some cosmological scenarios of interest. Although it is not a simple representation, studying high-dimensional gravity models through a mimic with testable gravitational scenarios seems an interesting path, since obtaining any observational data in high-dimensional gravity seems still far from being accomplished. In this sense, proposals such as $f(R)-$gravity, Gauss-Bonnet gravity, Palatini gravity, mimetic gravity, and so many others have been widely used in the study of branes \cite{Charmousis:2002rc,Kofinas:2003rz,Cho:2001nf,Afonso:2007gc,Bazeia:2008zx,Liu:2009ega,Borzou:2009gn,Zhong:2010ae,Bazeia:2013uva,Bazeia:2014poa,Gu:2014ssa,Bazeia:2015dna,Sui:2020atb}.

In the scope of generalized braneworld models, particular attention has been given to the study of situations that produce changes in the profile of the brane, in particular, the presence of internal structure in quantities such as energy density and/or warp factor. These modifications are important as they allow you to change the way gravity flows between four-dimensional space and hyperspace. In this perspective, proposals that modify Einstein's gravity such as Palatini gravity and $f(R,T)-$gravity have been described as good alternatives, since they introduce new degrees of freedom that can be adjusted accordingly \cite{Bazeia:2015owa,Correa:2015qma,Zhong:2017uhn,Zhong:2018fdq,Rosa:2022fhl}.

In addition to the mentioned proposals, recent studies have also shown that the inclusion of additional source fields with generalized dynamics can significantly alter the internal structure of the brane. In particular, cuscuton-like dynamics seem to develop an interesting game in this sense, since in different scenarios it has contributed to the emergence of some internal structure, modifying the graviton trapping profile in the brane \cite{Bazeia:2021jok,Rosa:2021myu,Bazeia:2022sgb}. We know that other proposals as, for example, mimetic gravity can also act to modify important qualities of the brane, producing a rich internal structure in the warp factor and energy density, as well as inducing modifications in the spectrum of the linear perturbations  \cite{Sadeghnezhad:2017hmr,Chen:2020zzs,Xiang:2020qrc}. In this sense, studying brane models considering together different modified scenarios also deserves some attention.

In this work, we focus on investigating brane in mimetic gravity, where the source fields have generalized dynamics and is organized as follows. In Sec. {\it 2} we describe the general formalism used for the study of brane with mimetic gravity. In this section, we also investigate specifics models that engender kink-like solutions. In Sec.~{\it 3} we analyze the linear stability of the model considering small perturbations in the fields. In Sec. {\it 4} we conclude and present perspectives for future work.

{\it{2. Thick brane in mimetic gravity.}} In this work, we study brane models in five-dimension spacetime with mimetic gravity where the source field has a non-canonical dynamic that is described in terms of an invariant $X=(1/2)\nabla_a\phi\nabla^a\phi$, where the Latin indices obey $a,b=1,2,\ldots,5$. The description that will be presented below starts from an action in the form
\begin{equation}\label{actiongeo}
S=\frac1{2\kappa}\int\!\sqrt{|g|}\Big(R-2\kappa{\cal L}_s(\phi,X)\Big)d^5x,
\end{equation}
where $\kappa$ is a positive coupling constant that must be positive, $g$ is the determinant of the metric tensor and ${\cal L}_s$ is the Lagrangian density of the source field that is written as 
\begin{equation}\label{lagrange}
{\cal L}_s(\phi,X)=\lambda\big(F(X)+U(\phi)\big)-V(\phi)\,.
\end{equation}
Here $\lambda$ is a scalar function that acts as a Lagrange multiplier. In the original mimetic model $F(X)=X$ and $U(\phi)=1$, however, in this paper we will consider more general situations. Although we can define an effective potential in terms of the Lagrange multiplier as $\bar{V}(\phi)=V(\phi)-\lambda U(\phi)$, we must take into account that $\lambda$ imposes a new constraint that should therefore add another equation into the system. Varying the action \eqref{actiongeo} concerning the source field we get
\ben\label{Eqsource}
\lambda\big(U_{\phi}-F_X\nabla_a\nabla^a\phi\big)-\nabla_a\big(\lambda F_X\big)\nabla^a\phi=V_{\phi},
\een
where $V_{\phi}=dV/d\phi$, $U_{\phi}=dU/d\phi$ and $F_X=dF/dX$. On the other hand, the Einstein equation has the usual form and can be written as
\ben\label{Eqeinste}
R_{ab}-\frac12g_{ab}R=\kappa T_{ab}\,.
\een
In addition to Eqs. \eqref{Eqsource} and \eqref{Eqeinste} we also obtain a constraint equation that comes from the variation of the action \eqref{actiongeo} concerning the Lagrange multiplier, that is,
\ben\label{Eqconstraint}
U(\phi)=-F(X)\,.
\een
By using the constraint equation given by \eqref{Eqconstraint} we can write the energy-momentum tensor $T_{ab}$ as
\ben
T_{ab}=\lambda F_X\nabla_a\phi\nabla_b\phi+g_{ab}V.
\een

As usually is done in the study of topological solutions in brane models, we consider static configurations for the fields. Thus, let us assume that the solution of the source field depends only on the extra dimension $y$, i.e., $\phi=\phi(y)$ and $\lambda=\lambda(y)$. Furthermore, we consider a metric in the form
\begin{equation}\label{metricbrane}
    ds_5^2=e^{2A}\eta_{\mu\nu}dx^\mu dx^\nu-dy^2\,,
\end{equation}
where the Greek indices $\mu,\nu$ run from $0$ to $4$ and the Minkowski metric $\eta_{\mu\nu}$ has a signature given by $(+,-,-,-)$. We also assume that the warp function is static, $A=A(y)$. In this way, we can write the equation of motion of the source field and the two non-vanishing independent components of the Einstein equation, respectively, as
\ben
\!\!\!\lambda \Big(4F_X A'\phi'\!+\!\left(F_X\!-\!F_{\text{XX}}\phi'^2\right)\!\phi''\!+\!U_{\phi}\!\Big)\!+\!F_X \lambda'\phi'\!=\!V_{\phi},\,\,\,\label{Eqmocstat}
\een
and
\bes
\bal
6A'^2=\,&\kappa\big(\lambda F_X \phi'^2- V(\phi)\big)\,,\label{compEin01}\\
3A''=\,&-\kappa\lambda F_X\phi'^2\,.\label{compEin02}
\eal
\ees
Here the prime represents the derivative concerning $y$. Note that we must deal with a set of four differential equations enumerated by Eqs. \eqref{Eqmocstat}, \eqref{compEin01}, \eqref{compEin02} and the constraint equation \eqref{Eqconstraint}. However, it is possible to show that only three of the differential equations are independent. For example, we can derive Eq. \eqref{compEin01} and use Eqs. \eqref{compEin02} and \eqref{Eqconstraint} to obtain the Eq. \eqref{Eqmocstat}.

To solve the system of equations described above, we  introduce the first-order formalism using two auxiliary functions $w(\phi)$ and $W(\phi)$ such that,
\ben\label{EqFO}
F_X\phi'=\frac{d w}{d \phi},\qquad\qquad A'=-\frac{\kappa}3W\,.
\een
The functions $w(\phi)$ and $W(\phi)$ are related through of the constraint equation in the form
\ben\label{constraEqs}
\frac{d W}{d \phi}=\lambda \frac{d w}{d\phi}.
\een
Note that the Lagrange multiplier can be written as a function of the field $\phi$. With this prescription, we can rewrite the potential $V$ as
\ben
V=\frac1{F_X}\frac{d w}{d \phi}\frac{d W}{d \phi}-\frac{2\kappa}3 W^2.
\een
Furthermore, the energy density that is obtained by $\rho(y)=-e^{2A}{\cal L}_s$, becomes
\ben\label{energdens}
\rho(y)=\frac{d}{dy}\big(e^{2A}W\big)\,.
\een

The study of scalar fields with modified dynamics has been investigated in several scenarios. To explore the formalism presented so far, we are going to choose some specific forms for $F(X)$, some of which have already been studied in brane, however, this is the first time they are used in mimetic gravity.

{\it{2.1. First case.}} The first model that we study is the kinematically modified case proposed in \cite{Bazeia:2008zx}, where the authors used a function $F(X)$ as
\ben\label{mod1}
F(X)=\frac{2^{n-1}}{n}X|X|^{n-1},
\een
where $n$ is a positive integer. Note that if $n=1$ we return to the usual dynamics commonly used to investigate braneworld models in mimetic gravity. By using static configurations, the constraint equation \eqref{Eqconstraint} assumes the form,
\ben
U(\phi)=\frac{1}{2n}\phi^{\prime 2n}.
\een
Furthermore, the first of the Eqs. \eqref{EqFO} becomes,
\ben\label{PpMod1}
\phi^{\prime 2n-1}=\frac{dw}{d\phi}.
\een
We can now choose $w(\phi)$ to get solutions for the source field. Let us follow Ref. \cite{Bazeia:2013euc} and consider
\be
w(\phi)=\phi\times\,{}_2F_1\Big(\frac12,1\!-\!2n;\frac32;\phi^2\Big)\label{wM1}\,,
\ee
where ${}_2F_1$ is the Hypergeometric function. In this case, the first-order equation for $\phi$ becomes $\phi'=1-\phi^2$, and the solution has the usual kink-like profile given by
\be\label{solm1}
\phi(y)=\frac{e^{2y}-1}{e^{2y}+1}.
\ee
See that the parameter $n$ does not change the profile of the solution. This behavior is driven by the choice of $w(\phi)$ in Eq. \eqref{wM1}. In general, the kink-like solution goes to asymptotic values $\phi_{a}$ when $|y|$ goes to infinity. For the solution \eqref{solm1} we obtain that $\phi_a=\pm1$. 

Let us consider now
\be
W(\phi)=\phi\,.\label{WM1}\\
\ee
Since the solution does not depend on $n$, the warp function, obtained by the second of Eqs. \eqref{EqFO}, will not depend on this parameter either. This means that all quantities that depend only on the warp function remain unaffected by the kinematic modification. For instance, the energy density, which is given by Eq.~\eqref{energdens} is written in terms of $A(y)$ and $W$, but $W$ in Eq. \eqref{WM1} only depends on $\phi$. So, the energy density does not depend on $n$. In Sec. {\it 3} we will also show that the stability becomes unchanged as it only depends on the warp function. This is an interesting characteristic of the model investigated in this section, where the kinematic modification does not affect the properties of the thick brane.

By using Eq. \eqref{constraEqs} and the definitions given by Eqs. \eqref{wM1} and \eqref{WM1}, we can write the Lagrange multiplier as $\lambda=(1-\phi^2)^{1-2n}$. Moreover, the potentials $U$ and $V$ can be written as
\bes
\bal
U(\phi)&=\frac1{2n}\big(1-\phi^2\big)^{2n},\label{potUM1}\\
V(\phi)&=1-\frac13\big(2\kappa+3\big)\phi^2\,.\label{potVM1}
\eal
\ees
When we use the asymptotic value of the solution in the potentials given by Eqs. \eqref{potUM1} and \eqref{potVM1} we obtain $U(\phi_a)=0$ and $V(\phi_a)=-2\kappa/3$. As we can see, the representation described leads to an asymptotically $AdS_5$ space, since $\bar{V}(\phi_a)\equiv \Lambda_5<0$.

\begin{figure}[ht]
    \begin{center}
        \includegraphics[scale=0.6]{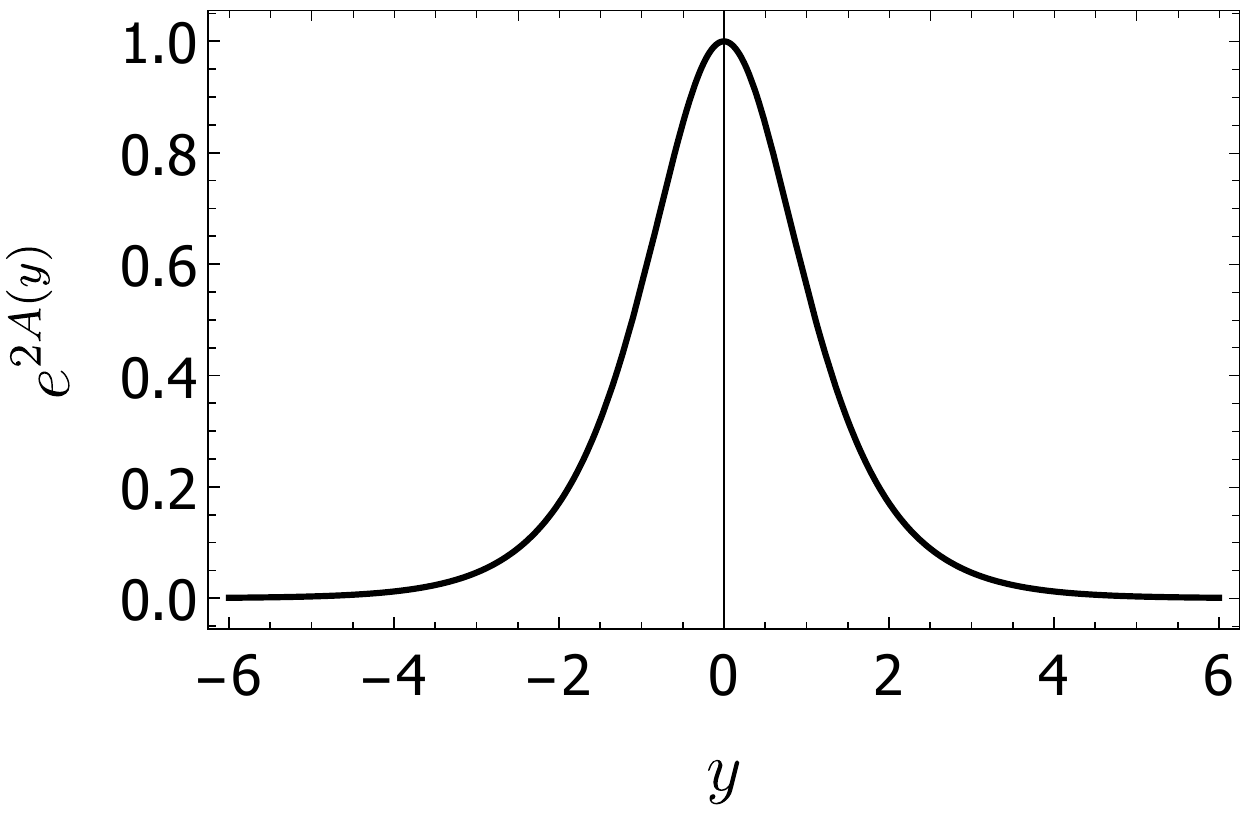}
        \includegraphics[scale=0.6]{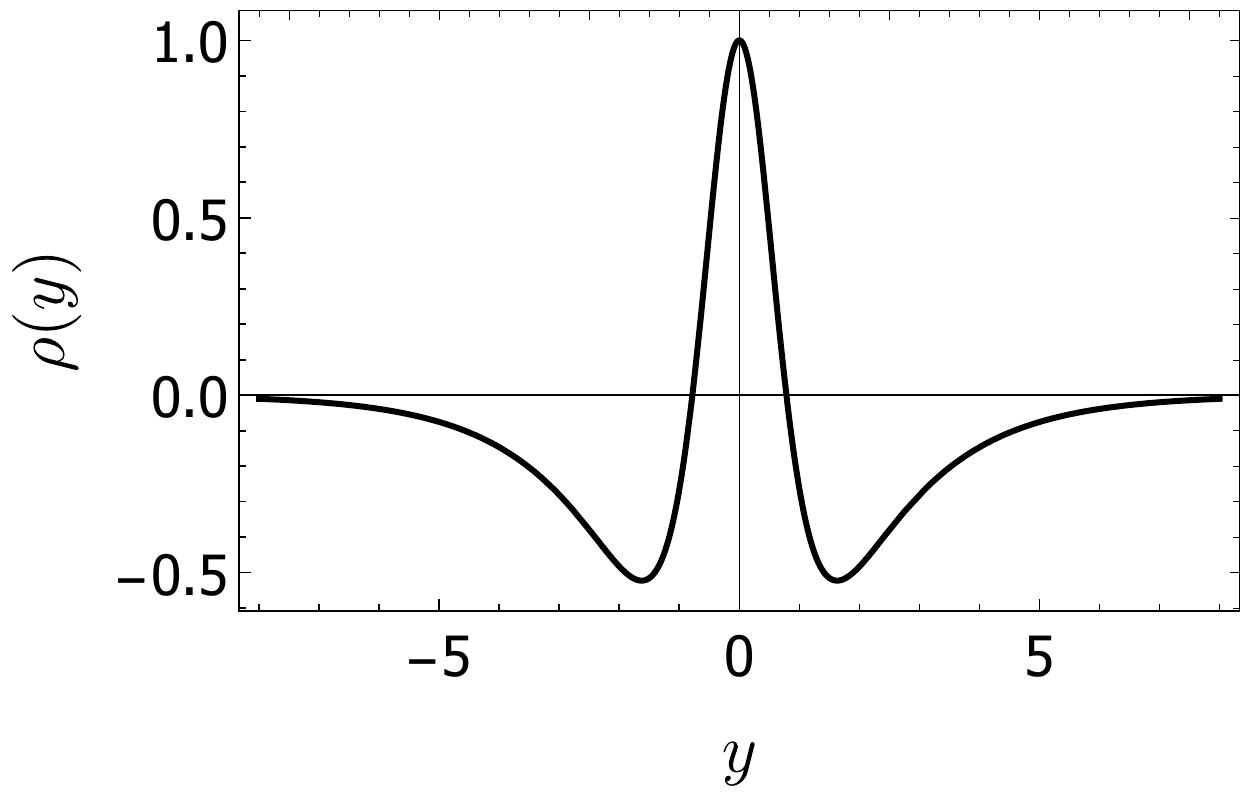}
    \end{center}
    \vspace{-0.5cm}
    \caption{\small{Warp factor (top panel) and energy density (bottom panel) depicted for $\kappa=2$.}\label{fig1}}
\end{figure}

Using the solution \eqref{solm1} we can write the Lagrange multiplied as $\lambda(y)=\cosh^{4n-2}(y)$. As we can see the Lagrange multiplier diverges asymptotically, however, this does not cause problems in the model since both the potential $U$ and the derivative of the solution go to zero. Thus, the term $\lambda(F(X)+U)$ in the action \eqref{actiongeo} is always well-behaved. To verify this, we calculate the warp factor and the energy density. By using the kink-like solution \eqref{solm1} in the second of Eqs. \eqref{EqFO} we obtain the warp function as
\be\label{warfunM1}
A(y)=\frac{\kappa}3\ln\big(\sech(y)\big).
\ee
As can be seen in top panel of Fig. \ref{fig1} the warp factor $e^{2A}$ has the usual bell shape profile. Furthermore, we obtain the thin brane profile as $A\approx-(\kappa/3) |y|$ when $|y|\gg 0$. The energy density can be found by Eq. \eqref{energdens}, being written as
\be\label{enerdenM1}
\rho(y)=\sech^{\kappa/3}(y)\left(\Big(\frac{2 \kappa}{3}+1\Big) \sech^2(y)-\frac{2\kappa}{3}\right).
\ee
In bottom panel of Fig. \ref{fig1} we show the energy density obtained by Eq. \eqref{enerdenM1} for $\kappa=2$. As we can see, everything works in the usual way and no unwanted behavior appears. We also verified the behavior of the Ricci scalar $R$ and the Kretschmann scalar $K=R_{abcd}R^{abcd}$, where $R_{abcd}$ is the Riemann tensor, and everything seems to work properly.

Another possibility also investigated in Ref. \cite{Bazeia:2008zx} enables compact solutions. To examine this new situation, let us assume that $w(\phi)$ has the form
\be
w(\phi)=\phi\times\,{}_2F_1\Big(\frac12,\frac12\!-\!n;\frac32;\phi^2\Big)B_1(\phi)+B_2(\phi),\label{wM11}
\ee
where
\be\nonumber
B_1(\phi)=
\begin{cases}
i(-1)^n,& \mbox{for}\quad  \phi^2>1  \\
1, & \mbox{for}\quad \phi^2\leq1
\end{cases},
\ee
\be\nonumber
B_2(\phi)=
\begin{cases}
\!\big(1\!-\!i(-1)^n\big)\frac{\sqrt{\pi}\,\Gamma(n+1/2)}{\Gamma(n+1)}\sgn(\phi),& \!\!\mbox{for}\!\quad  \phi^2>1  \\
0, & \!\!\mbox{for}\!\quad \phi^2\leq1
\end{cases},
\ee
and $\sgn$ represent the signum function and $\Gamma$ is the gamma function defined by
\ben
\Gamma(s)=\int_{0}^{\infty}p^{s-1}e^{-p}dp\,.
\een
In this case, the source field solution is written in the form
\be
\phi(y)=
\begin{cases}
\sgn(y),& \mbox{for}\quad  |y|>\pi/2  \\
\sin(y),& \mbox{for}\quad  |y|\leq \pi/2
\end{cases}.
\ee
Again, we have chosen $w(\phi)$ so that the parameter $n$ does not influence the main characteristics of the model. If we consider that $W$ is still given by Eq. \eqref{WM1}, we obtain the warp function as
\be\label{warM1com}
A(y)\!=\!
\begin{cases}
-(\kappa/3)|y|\!+\!(\kappa/6) (\pi\!-\!2),& \!\!\mbox{for}\!\quad  |y|>\pi/2  \\
-(\kappa/3)\big(1\!-\!\cos(y)\big),& \!\!\mbox{for}\!\quad  |y|\leq \pi/2
\end{cases},
\ee
where we considered $A(0)=0$ and enforced the continuity of the warp function at $y=\pm\pi/2$. Note that the brane becomes hybrid, that is, for $|y|\leq\pi/2$ we have a thick brane, but for $|y|>\pi/2$ we have a thin brane. The warp function can be used to write the energy density as
\be\label{rhoM1com}
\!\!\rho(y)\!=\!
\begin{cases}
-(2\kappa/3)e^{\kappa(\pi-2)/3-2\kappa|y|/3},& \!\!\mbox{for}\!\quad  |y|>\pi/2  \\
h(y)\,e^{-(2\kappa/3)(1-\cos(y))},& \!\!\mbox{for}\!\quad  |y|\leq \pi/2
\end{cases},
\ee
where $h(y)=\cos(y)\!-\!(2\kappa/3)\sin^2(y)$. The behavior of the warp factor and energy density given by Eqs. \eqref{warM1com} and \eqref{rhoM1com} are similar to the case shown in Fig. \ref{fig1}, so we omit these figures here.

{\it{2.2. Second case.}} Let us consider a model with cuscuton-like dynamics in the form
\ben\label{Fcuscuton}
F(X)=X+\alpha \sqrt{|2X|}\,,
\een
where $\alpha$ is a positive parameter that controls the cuscuton term. Different from the previous model, the change in dynamics will work together with the usual dynamics. To understand the role of the cuscuton parameter in mimetic gravity we will keep the form of $W$ given by Eq.~\eqref{WM1} and consider only the case where $n=1$ in the Eq.~\eqref{wM1} that is,
\ben
w(\phi)=\phi-\frac13\phi^3\,.
\een
In this case, the Lagrange multiplier is $\lambda=1/(1-\phi^2)$ and the potentials $U$ and $V$ becomes
\bes
\bal
U(\phi)&=\frac{1}{2} \left(1-\alpha-\phi^2\right) \left(1-3\alpha-\phi^2\right),\label{potUM2}\\
V(\phi)&=1+\alpha-\frac13(2\kappa+3)\phi^2+\frac{2\alpha ^2}{1-2\alpha-\phi^2} \,.\label{potVM2}
\eal
\ees
The first-order equation for the source field is
\ben\label{PpMod3}
\phi^{\prime}=1+\alpha-\phi^2.
\een
Again the solution has a kink-like profile and can be written as
\be\label{solcuscuton}
\phi(y)=\sqrt{1\!+\!\alpha}\,\tanh\big(\sqrt{1\!+\!\alpha}\,y\big).
\ee
Note that here the cuscuton parameter controls the thickness of the solution and changes the asymptotic value which now is $\phi_a=\pm\sqrt{1\!+\!\alpha}$ when $y\to\pm\infty$. With that we get
\be
U(\phi_a)=4\alpha^2\,,\quad V(\phi_a)=-\frac{2}{3}\big(\kappa+\alpha(1 +\kappa)\big).
\ee

\begin{figure}[ht]
    \begin{center}
        \includegraphics[scale=0.6]{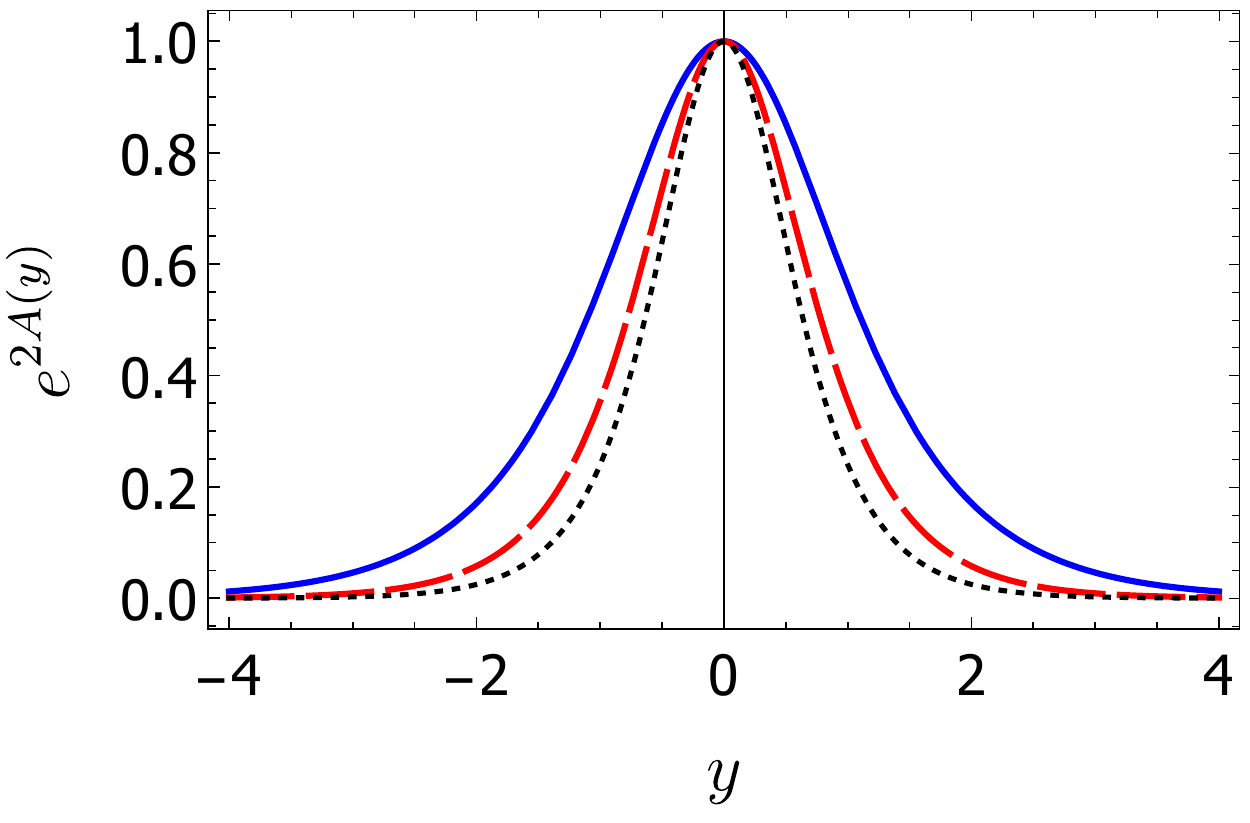}
        \includegraphics[scale=0.6]{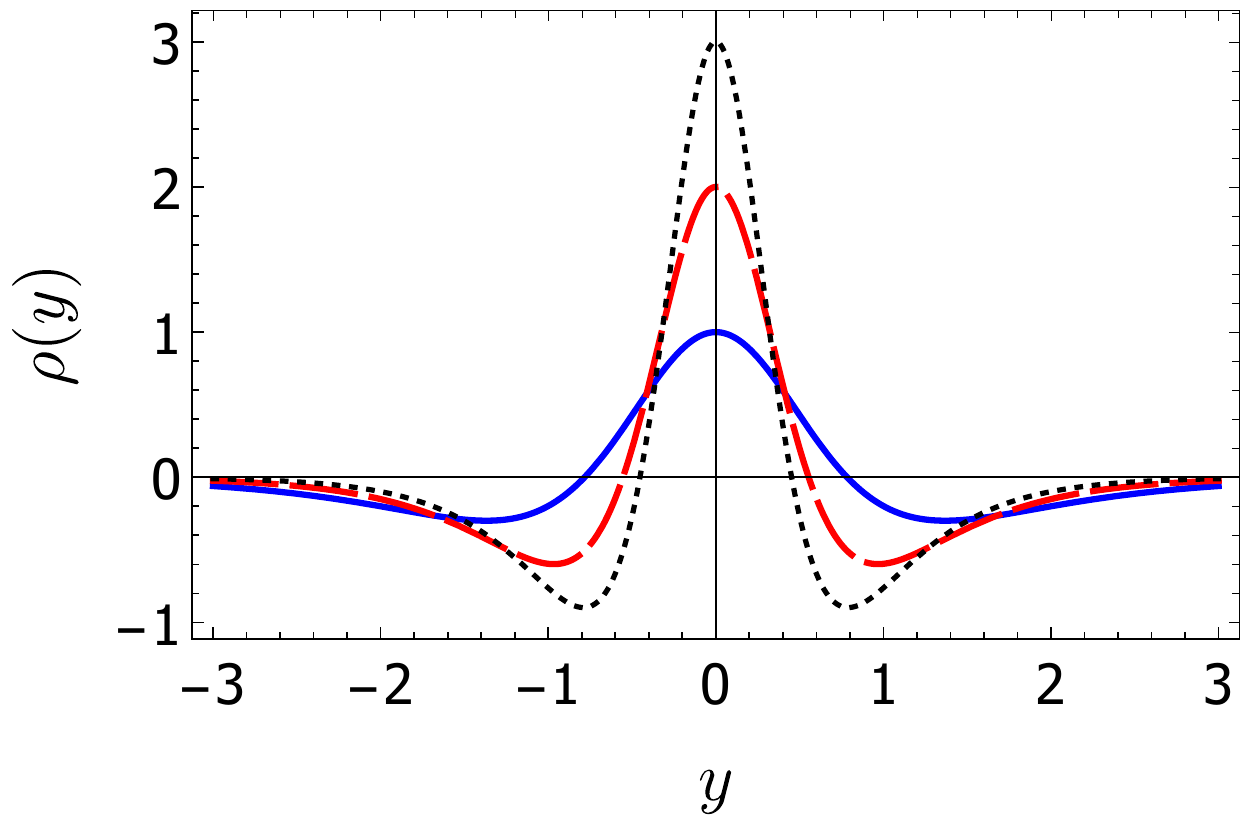}
    \end{center}
    \vspace{-0.5cm}
    \caption{\small{Warp factor (top panel) and energy density (bottom panel) depicted for $\kappa=2$ and $\alpha=0$ (solid-line), $\alpha=1$ (dashed-line) and $\alpha=2$ (dotted-line).}\label{fig2}}
\end{figure}

We obtain the warp function and the energy density, respectively, as
\be
A(y)=\frac{\kappa}3\ln\left(\sech\big(\sqrt{\alpha\!+\!1}\,y\big)\right),
\ee
\ben
\rho(y)&=&(1+\alpha)\,\sech^{2\kappa/3}\big(\sqrt{\alpha\!+\!1}\,y\big)\nn
&&\times\left(\Big(\frac{2\kappa}3+1\Big)\,\sech^{2}\big(\sqrt{\alpha\!+\!1}\,y\big)-\frac{2\kappa}{3}\right).
\een
In the Fig. \ref{fig2} we represent the energy density and the warp factor for $\kappa=2$ and $\alpha=0,1$ and $2$. As we can see, the cuscuton parameter modifies the characteristic width of these quantities around the origin. However, the behavior is typically the same as in the case without cuscuton. In this case, the thin brane profile is obtained as $A(y)\!\approx\!-(\kappa/3)\sqrt{1\!+\!\alpha}\,|y|$ when $|y|\!\gg\!0$.

{\it{3. Linear Stability.}} Although the theory defined by action \eqref{actiongeo} has introduced a new scalar degree of freedom through the Lagrange multiplier, we can verify that the model is stable for linear perturbations in the fields. In general, this analysis is performed considering small perturbations in the scalar field and the metric tensor in the form $\phi\to\phi(y)+\delta\phi(r,y)$ and $g_{ab}\to g_{ab}(y)+\pi_{ab}(r,y)$, where $r=x^\mu$ is the four-dimensional position vector. Furthermore, the four-dimensional part of $\pi_{ab}$ satisfies the transverse and traceless (TT) conditions, i.e., $\partial^{\mu} \pi_{\mu\nu}=0$ and $\pi_{\mu}{}^\mu=0$.

Let us also consider that $\pi_{\mu\nu}=e^{2A(y)}h_{\mu\nu}(r,y)$. Substituting the perturbation of the fields in Eq. \eqref{Eqeinste} and considering the static Eqs. \eqref{compEin01} and \eqref{compEin02}, we obtain the equation for $h_{\mu\nu}$ as
\begin{equation}\label{eqstabili}
\left(\partial_y^2+4A^\prime \partial_y \right)h_{\mu\nu}=e^{-2A}\partial_\alpha\partial^\alpha h_{\mu\nu} \,.
\end{equation}
If we consider a new $z$-coordinate as $dz=e^{-A(y)}dy$ and also if $h_{\mu\nu}(r,z)=H_{\mu\nu}( z)\,e^{-3A(z)/2}\,e^{-ik\cdot r}$, we can rewrite the Eq. \eqref{eqstabili} in the form
\ben\label{schrodinger}
-\frac{d^2H_{\mu\nu}}{dz^2}+{\cal U}(z)H_{\mu\nu}= k^2H_{\mu\nu}\,,
\een
where
\ben\label{potschrodinger}
{\cal U}(z)=\frac32A_{zz}+\frac94A_z^2\,.
\een
The Eq. \eqref{schrodinger} can be factorized as $S^{\cal y} S\,H_{\mu\nu}= k^2H_{\mu\nu}$, where $S^{\cal y}=-\partial_z+(3/2)A_z$ and $k^2\geq 0$, indicating that the system is stable for linear perturbations. To represent the quantities involved in stability we can rewrite the potential \eqref{potschrodinger} in the variable $y$ as
\be\label{potschrodinger2}
{\cal U}(y)=\frac34 e^{2A}\!\left(5 A'^2+2A''\right).
\ee
Furthermore, we can also represent the zero mode $\xi(y)$ as
\ben\label{zeromode}
\xi(y)=Ne^{2A(y)},
\een
where $N$ is a normalization constant. In a direct way, we can find the complete expressions of the stability potential and zero mode for the two models studied in this work.

\begin{figure}[t]
    \begin{center}
        \includegraphics[scale=0.6]{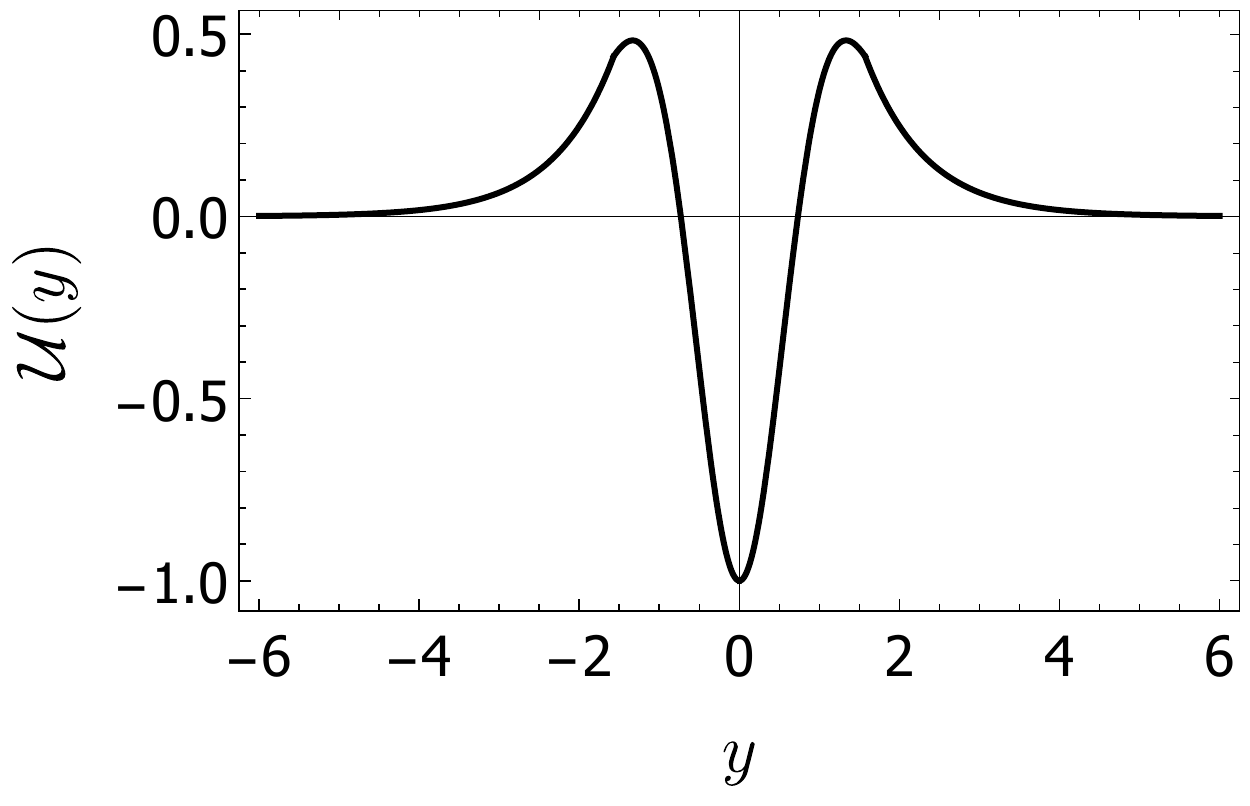}
        \includegraphics[scale=0.6]{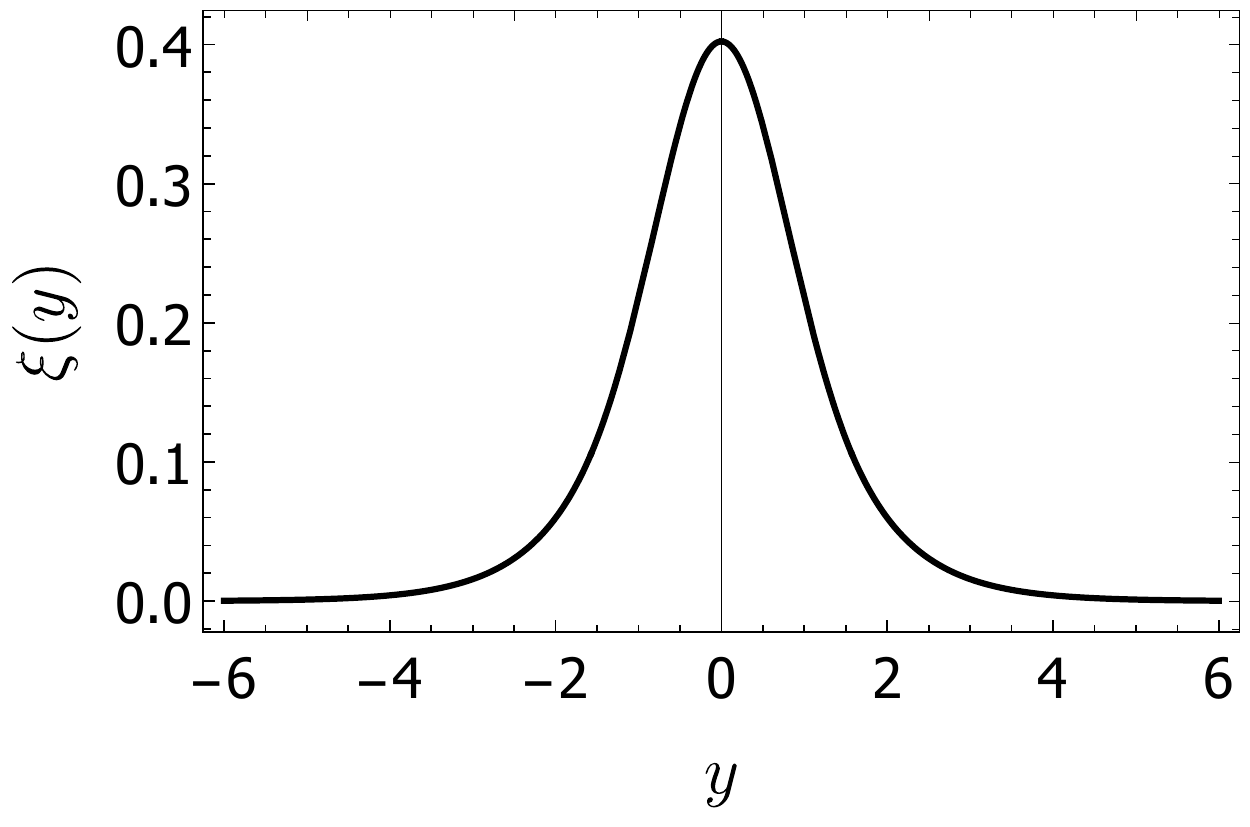}
    \end{center}
    \vspace{-0.5cm}
    \caption{\small{Stability potential (top panel) and zero mode (bottom panel) depicted for compact solution with $\kappa=2$.}\label{fig3}}
\end{figure}

\begin{figure}[t]
    \begin{center}
        \includegraphics[scale=0.6]{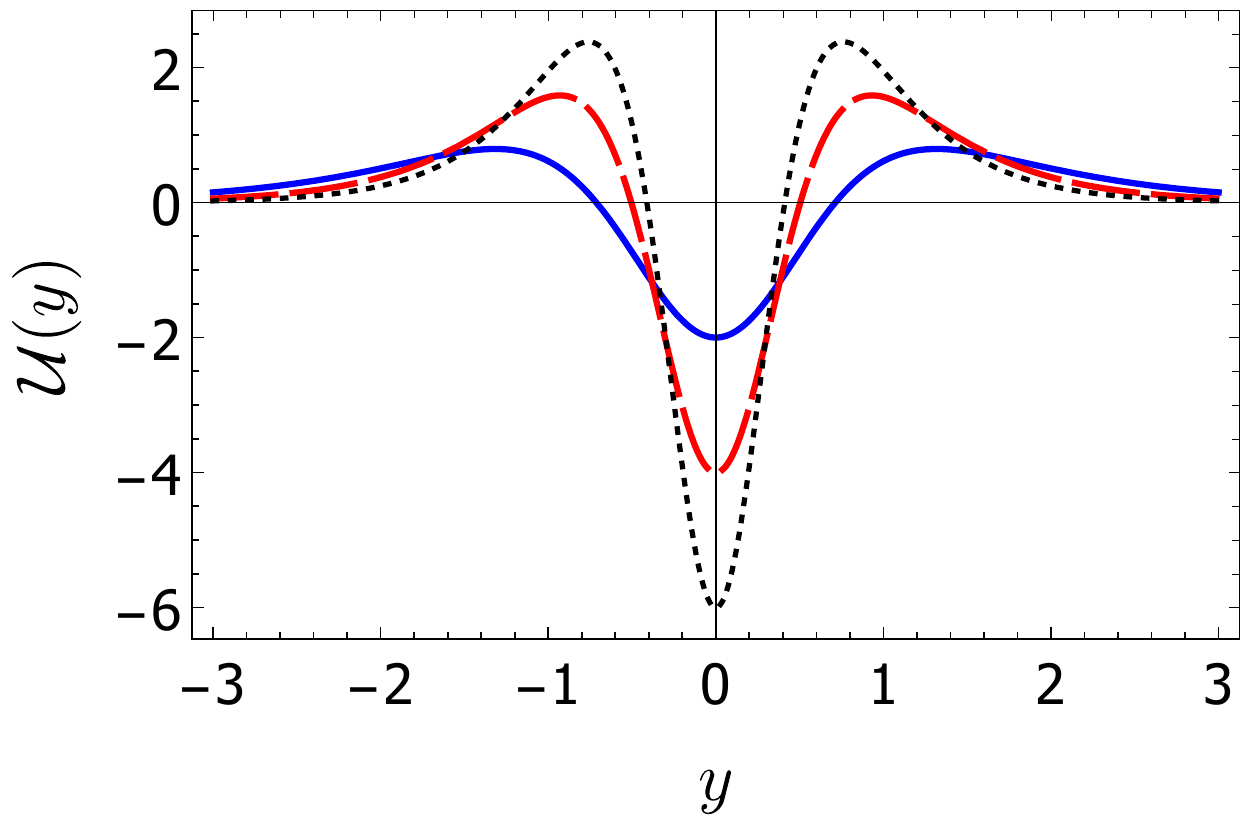}
        \includegraphics[scale=0.6]{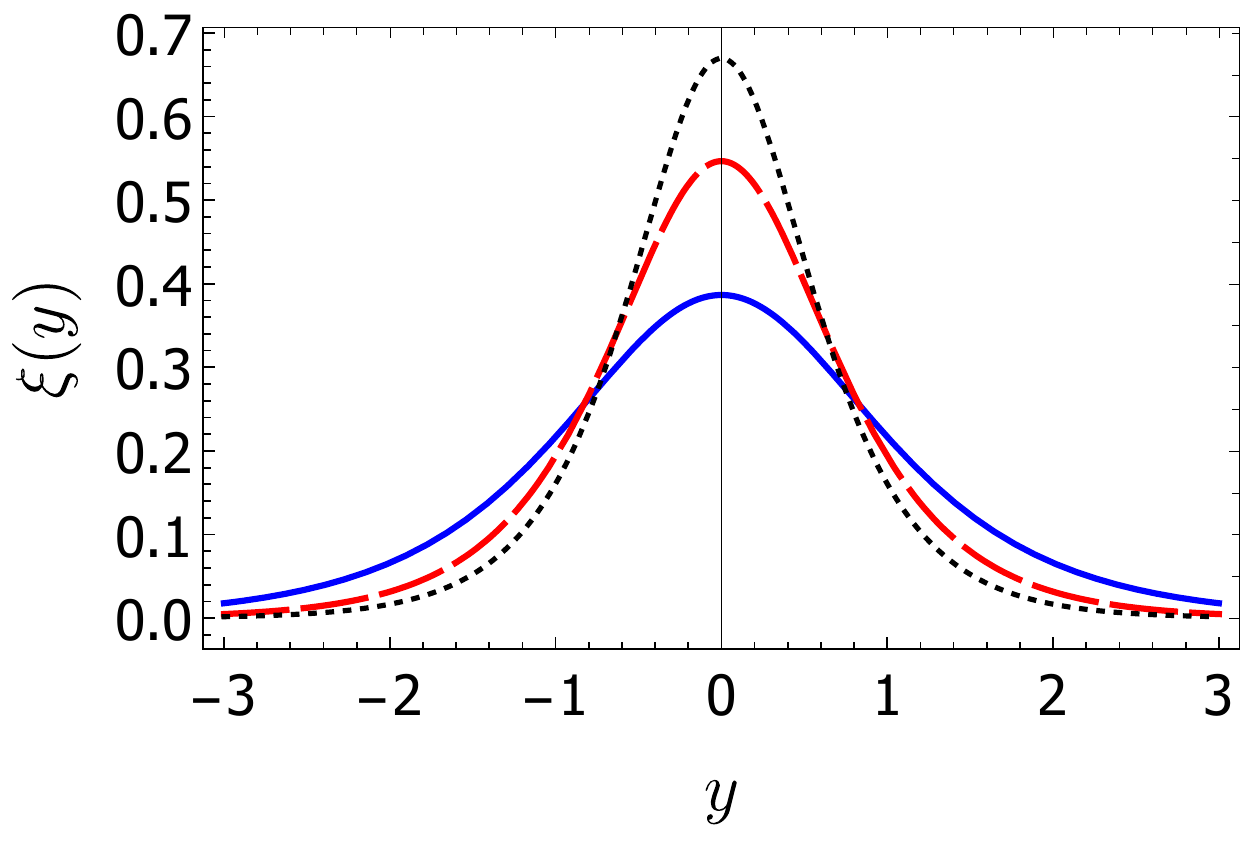}
    \end{center}
    \vspace{-0.5cm}
    \caption{\small{Stability potential (top panel) and zero mode (bottom panel) depicted for $\kappa=2$ and $\alpha=0$ (solid-line), $\alpha=1$ (dashed-line) and $\alpha=2$ (dotted-line).}\label{fig4}}
\end{figure}

For the first model with the solution studied in section {\it 2.1}, we have
\bes\label{eqstabiliM1a}
\bal
{\cal U}(y)&=\frac{\kappa}{6}\, \sech^{2\kappa/3}(y)\Big(5 \kappa\!-\!(5\kappa\!+\!6)\, \sech^2(y)\Big)\,,\label{estam1}\\
\xi(y)&=\frac{\Gamma \big(\frac{\kappa}{3}\!+\!\frac12\big)}{\sqrt{\pi}\, \Gamma \big(\frac{\kappa}{3}\big)}\,\sech^{2\kappa/3}(y)\,.\label{modzerom1}
\eal
\ees
As we commented before, the stability from Eq. \eqref{potschrodinger2} and the zero mode from \eqref{zeromode} only depend on the warp factor. So, both ${\cal U}(y)$ and $\xi(y)$ do not depend on $n$, as shown in \eqref{eqstabiliM1a}. The solid line in Fig. \ref{fig4} shows the behavior of the stability potential and zero mode given by Eqs. \eqref{eqstabiliM1a}, as we can see the stability potential maintains the Vulcan profile, and the zero mode is located in the extra dimension. Now, for the compact solution, we get,
\be
\!\!{\cal U}(y)\!=\!
\begin{cases}
\!(5\kappa^2/12)e^{\kappa(\pi-2)/3-2\kappa|y|/3},\!\!& \!\!\mbox{for}\,  |y|\!>\!\pi/2  \\
\!(\kappa/2)p(y)e^{-(2\kappa/3)(1-\cos(y))},\!\!& \!\!\mbox{for}\,  |y|\!\leq\! \pi/2
\end{cases}\!,
\ee
where $p(y)=(5\kappa/6)\sin^2(y)\!-\!\cos(y)$, and
\be
\!\!\xi(y)\!=\!
\begin{cases}
\!Ne^{\kappa(\pi-2)/3-2\kappa|y|/3},\!\!& \!\!\mbox{for}\,  |y|\!>\!\pi/2  \\
\!Ne^{-(2\kappa/3)(1-\cos(y))},\!\!& \!\!\mbox{for}\,  |y|\!\leq\! \pi/2
\end{cases}\!,
\ee
where the normalization constant $N$ is
\be
N=\kappa e^{2\kappa/3}\left(\pi  \kappa {L}_0\left(\frac{2 \kappa}{3}\right)+\pi  k I_0\left(\frac{2 k}{3}\right)+3\right)^{-1},
\ee
where $I_0$ and ${L}_{0}$ are, respectively, the modified Bessel and Struve functions, defined as
\ben
I_n(z)=\frac1\pi\int_0^\pi e^{z\cos\theta}\cos(n\theta)d\theta\,,
\een
and
\ben
L_n(z)=\frac{z^{n+1}}{2^n\sqrt{\pi}\Gamma\big(n+\frac32\big)}\,{}_1F_2\Big(1;\frac32,n+\frac32;-\frac{z^2}4\Big).
\een
Fig. \ref{fig3} shows the stability potential and zero mode for the solution with a compact profile. Here again, everything works as it should.

For the model with cuscuton dynamics studied in section {\it 2.2} we obtain
\bes
\bal
{\cal U}(y)=&\,\frac{\kappa}{6}(1\!+\!\alpha)\, \sech^{2\kappa/3}\big(\sqrt{1\!+\!\alpha}\,y\big)\nn
&\,\times\Big(5 \kappa\!-\!(5\kappa\!+\!6)\, \sech^2\big(\sqrt{1\!+\!\alpha}\,y\big)\Big)\,,\label{estaM2}\\
\xi(y)=&\,\sqrt{1\!+\!\alpha}\,\frac{\Gamma \left(\frac{\kappa}{3}\!+\!\frac{1}{2}\right)}{\sqrt{\pi}\, \Gamma \left(\frac{\kappa}{3}\right)}\,\sech^{2\kappa/3}\big(\sqrt{1\!+\!\alpha}\,y\big)\label{ModzerM2}\,.
\eal
\ees
Note that if $\alpha=0$ the previous equations reproduce the Eqs. \eqref{estam1} and \eqref{modzerom1}. In top panel of Fig. \ref{fig4} we show the stability potential given by Eq. \eqref{estaM2} for some values of the cuscuton parameter. In bottom panel of the same figure, we show the zero mode given by Eq. \eqref{ModzerM2}. In both cases we use $\kappa=2$ and $\alpha=0,1$ and $2$. Note that the cuscuton parameter changes the depth of the stability potential and concentrates the zero mode near the origin.

{\it{4. Conclusions.}} In this work we have studied brane models in generalized mimetic gravity, assuming that the source field has non-canonical dynamics. We started the study by describing the formalism needed to investigate brane models in generalized gravity, where we obtained the equations of the fields and the new constraint equation that arises from the variation of the action concerning the Lagrange multiplier. To obtain topological solutions, we considered static configurations for the fields and solved the equations of motion using first-order formalism. Next, we investigated two specific dynamics that allow analytical solutions, the first being a case of kinematic modification proposed in \cite{Bazeia:2008zx} and the second a cuscuton-like dynamic. Finally, we also investigated linear stability on general grounds and for each type of dynamics studied.

In the first model, we considered two types of kink-like solutions. The first had a standard profile in the form $\tanh(y)$ and the second had a compact profile. For the standard kink solution, we found that the brane behaved in a usual way, with the warp factor and energy density profile similar to those found in other brane models. On the other hand, when the source field solution has a compact profile, the brane becomes hybrid, having a distinct behavior inside and outside of the compact space.

In the second model, we introduced a cuscuton-like dynamic which was controlled by a real positive parameter. For simplicity, we only investigated the case of the usual kink solution. We realized that the cuscuton parameter only changes the thickness of the brane, modifying the distribution of the warp factor and the energy density around the origin. However, the quantities investigated have essentially the same profile as in the case without the cuscuton contribution.

As can be seen in this work, it is possible to build brane models in mimetic gravity, with the fields having unusual dynamics. We were able to find analytical solutions in different situations and the theory is stable by linear perturbations. This indicates that the generalization proposed in this work is robust and can be introduced to change the behavior of the brane.

As a continuation of this proposal, other scalar fields could be added as sources of the brane. We have investigated that the addition of new scalar fields, together with a cuscuton-like dynamic, can significantly alter the internal structure of the brane \cite{Bazeia:2021jok,Rosa:2021myu,Bazeia:2022sgb}. This seems to be promising, having led to distinct and interesting changes in the standard brane scenario. Furthermore, studying mimetic gravity in alternative representations as, for example, tensor-scalar representation of $f(R,T)-$gravity, can also induce new behaviors that are not found in the usual scenario. Some of these issues are currently under investigation and we hope to report on them in the near future.

{\it Declaration of competing interest:} The author declares that he has no known competing financial interests or personal relationships that could have influenced the results of the work.

{\it Data availability:} This manuscript describes a theoretical work with no associated data to be deposited.

{\it Acknowledgments:} The author would like to thank Dionisio Bazeia for the discussions. This work is partially financed by Para\'iba State Research Foundation, FAPESQ-PB, grant No. 0015/2019.


\end{document}